\documentclass[pra,showpacs,superscriptaddress,amssymb,twocolumn,longbibliography,nofootinbib]{revtex4-1} 
\usepackage[utf8]{inputenc}

\usepackage{float}
\usepackage{amsmath,amssymb,setspace}
\usepackage{graphicx}
\usepackage{subfigure}  
\usepackage[usenames]{color}
\usepackage{graphicx}
\usepackage{bm, bbm}      
\usepackage{color}
\usepackage[dvipsnames]{xcolor}
\usepackage{srcltx}
\usepackage{upgreek}
\usepackage{slashed}
\usepackage{dirtytalk}
\usepackage{mathtools}
\usepackage{titlesec}
\usepackage{titlecaps}

\usepackage{algorithmicx}
\usepackage[ruled,vlined]{algorithm2e}
\SetAlgoCaptionSeparator{ }

\usepackage{xcolor}

\usepackage{bm, bbm}      
\usepackage{color}
\usepackage{srcltx}

\usepackage{slashed}

\def\=d{\, {\buildrel \rm def  \over =} \,}

\def\sqr#1#2{{\vcenter{\vbox{\hrule height.#2pt \hbox{\vrule width.#2pt height#1pt \kern#1pt \vrule width.#2pt}\hrule height.#2pt}}}}

\def\beq#1{\begin{equation} \label{#1}}
\def\eeq{\end{equation}}
\def\ben{\begin{equation*}}
\def\een{\end{equation*}}
\def\bequa{\begin{eqnarray}}
\def\eequa{\end{eqnarray}}

\def\bf#1{\bm{#1}}

\newcommand\bea{\begin{eqnarray}}
\newcommand\eea{\end{eqnarray}}

\def\beq{\begin{equation}}
\def\eeq{\end{equation}}

\def\al{\alpha}

\newcommand{\hini}{H_{\text{ini}}}

\newcommand{\hnav}{H_{\text{nav}}}
\newcommand{\hfin}{H_{\text{fin}}}
\newcommand{\ang}{\overset{\circ}{\text{A}}}

\newcommand{\vtd}{VSQS}
\newcommand{\vrs}{VSQS}

\begin{document}

\title{Variationally Scheduled  Quantum Simulation}

\author{Shunji Matsuura}
\email{shunji.matsuura@1qbit.com}
\affiliation{1QB Information Technologies (1QBit) \\200-1285 Pender St W, Vancouver, BC, Canada}

\author{Samantha Buck}
\affiliation{1QB Information Technologies (1QBit) \\200-1285 Pender St W, Vancouver, BC, Canada}
\affiliation{Department of Physics, University of Guelph  \\50 Stone Rd E, Guelph, ON, Canada}

\author{Valentin Senicourt}
\affiliation{1QB Information Technologies (1QBit) \\200-1285 Pender St W, Vancouver, BC, Canada}

\author{Arman Zaribafiyan}
\affiliation{1QB Information Technologies (1QBit) \\200-1285 Pender St W, Vancouver, BC, Canada}

\begin{abstract}
    \begin{center}
        \vspace{-.9em}
        (Date: \today)
    \end{center}

Eigenstate preparation is ubiquitous in quantum computing, and a standard approach for generating the lowest-energy states of a given system is by employing adiabatic state preparation~(ASP).
In the present work, we investigate a variational method for determining the optimal scheduling procedure within the context of ASP.
In the absence of quantum error correction, running a quantum device for any meaningful amount of time causes a system to become susceptible to the loss of relevant information.
Therefore, if accurate
quantum states are to be successfully generated, it is crucial to find techniques that shorten the time of individual runs during iterations of annealing. We demonstrate our variational method toward this end by investigating the hydrogen and P4 molecules, as well as the Ising model problem on a two-dimensional triangular lattice. In both cases, the time required for one iteration to produce accurate results is reduced by several orders of magnitude in comparison to what is achievable via standard ASP. 
The significant shortening of the required time is achieved by using excited states partially during ASP.
As a result, the required quantum coherence time to perform such a calculation on a quantum device becomes much less stringent with the implementation of this algorithm. In addition, our variational method is found to exhibit resilience against control errors, which are commonly encountered within the realm of quantum computing.
\end{abstract}

\maketitle

\titleformat{\section}
  {\large\bfseries\center}{\thesection.}{10pt}{\MakeUppercase}{}

\section{Introduction}

It is widely recognized that attempts to realize solutions to computational problem sets involving quantum systems via classical hardware quickly give rise to intractable bottlenecks. This consequence becomes readily apparent as  the  amount  of  parameters  required  to  describe  the quantum state in question increases exponentially with growing system size.  A commonly encountered manifestation of this phenomenon presents itself within the context of quantum chemistry, where issues such as improving the computational efficiency of determining electronic correlation energies are an ongoing endeavour~\cite{meiwes2000metal,pilar2001elementary}.

Solutions to problems of this nature are highly desirable, yet finding them is notoriously
computationally challenging, and far beyond the capabilities of
even the most powerful of present-day supercomputers.
Researchers have tried to mitigate this issue of intractability by implementing heuristics and approximation techniques such as density functional theory (DFT) 
in attempts to decrease the computational cost of simulating more complex molecular systems~\cite{cohen2008insights,dobson2013electronic}. Nevertheless, even state-of-the-art approaches 
such as the employment of coupled-cluster techniques are met with considerable limitations, as they can accurately handle, \textit{at most}, molecules of a few dozen atoms in size~\cite{jeziorski1981coupled,kvasnivcka1982coupled, bartlett2007coupled}. 
Essentially, the common obstacle that these techniques all share can be understood as an established trade-off between the computational efficiency of the solver, and the accuracy of the obtained approximate solution.

As a promising initiative to alleviate this issue and reduce the   resources required, there has been increasing interest in solving quantum problems using quantum devices~\cite{Martinis9qubit,Bernien:2017aa,IonQ-11qubits}.
These methods stand to offer a more scalable alternative that would help circumvent the limitations currently imposed by classical computation.
 However, the main challenge faced in the near term for quantum devices is the absence of error correction techniques~\cite{PreskillNISQ};
without proper error corrections, computational results may not be reliable due to qubit decoherence and control errors.
Two ways to overcome drawbacks of this nature that present themselves in the absence of error correction is therefore to shorten the time taken to perform a single run of a quantum algorithm, and to make the computational process as noise resilient as possible.
One way to achieve these aims is by
employing variational methods~\cite{PeruzzoPhotonicPQ,VQE2,VQE3}
using
shallow circuit ansatz (e.g., see~\cite{Hardware-efficientIBM, 2018arXiv180101053D,PastCausalCones}) in quantum--classical hybrid algorithms.

In this paper, we consider the adiabatic approach of quantum computation for solving the problem of finding eigenstates.
Adiabatic state preparation as a computational approach offers many desirable features, 
such as robustness against various types of noise~\cite{Albash:2015nx,childs_robustness_2001,PhysRevLett.95.250503,quant-ph/0507010,PhysRevA.71.032330},
the absence of Trotterization errors, and the absence of accuracy limitations that arise from the requirement of having highly complicated ansatz.
Despite these advantages, there remain  factors that influence the choice of the annealing time $T$ that can affect the accuracy of the obtained results.

In ASP, if $T$ is defined to be too small, one subjects the system to the potential of undergoing harmful non-adiabatic transitions, and as a result the computation becomes susceptible to finding inaccurate results.  
Conversely, selecting too large a value of $T$ can result in the loss of quantum information due to decoherence.
The major challenge of near-term devices is that decoherence may occur quite early---before the adiabaticity condition is met---the consequence of which is that we obtain an inaccurate result. 

Therefore, the main objective of this work is to develop an algorithm in which $T$ can be chosen small enough to avoid decoherence, while simultaneously avoiding 
harmful types of non-adiabatic transitions, which can be understood as the specific non-adiabatic transitions that prevent quantum states from reaching the true ground state at
the end of the computation.
Based on previous works~\cite{VanQver,VQE3},
we consider a variational method to achieve this 
objective.

In the context of molecular systems, it was  successfully demonstrated
that a significant reduction in the required annealing time per individual run
was achieve to a degree of  
accuracy within a certain threshold, compared to the standard ASP method~\cite{VanQver}.

This was achieved 
by initializing a set of new terms, denoted the ``navigator Hamiltonian'', during the annealing schedule. 
Each respective Hamiltonian involved in the computation---the initial Hamiltonian,  navigator Hamiltonian, and final Hamiltonian---was assigned a predetermined scheduling function, similar to the settings employed in~\cite{Farhi:differentPaths,Perdomo-Ortiz:2011,crosson2014different,VQE3,PhysRevResearch.2.013283}.
Additionally, the coefficients of the defining terms present in both the navigator Hamiltonian and the initial Hamiltonian were defined variationally. 
It is worth noting that in the case of molecular systems investigated in~\cite{VanQver}, the reduction in annealing time required to achieve a 
specified accuracy was enabled in part by accessing excited states during annealing.

A natural consequence of utilizing this variational technique towards establishing the schedule functions is that the accuracy of the obtained quantum state 
is highly dependent on the schedule functions of the Hamiltonians involved.
Therefore, finding optimal schedule functions is essential in investigating whether quantum annealing can provide an advantage in solving combinatorial optimization problems.
Various forms of schedule functions have been considered both theoretically and experimentally.
Examples include the inhomogeneous transverse field~\cite{InhomogeneousMcMahon,Inhomo-DicksonAmin,InhomoDriver,Inhomo-pspin}, which is characterized by individual qubits possessing distinct transverse field strengths;
the anneal ``pause and quench'' schedule~\cite{PowerofPausing,Pause.p-spin}, where the scheduling functions are held constant for a certain period of time followed by a rapid modification;
and reverse annealing~\cite{Perdomo-Ortiz:2011,Chancellor:2016ys,DwaveReversAnnealTopological,ReverseAnnealLowRank,ReverseAnnealPortfolio,ReverseAnnealingONL,DynamicsofReverseAnneal,PhysRevA.101.022331},
which operates by starting a system in a classical state, introduces quantum fluctuations, and finally terminates by the removal of these fluctuations.\\

It is also worth mentioning that QAOA is a specific realization of  variationally scheduled quantum simulation (VSQS). 
QAOA specifically caters to combinatorial optimization problems and employs a discrete optimization technique.
It applies an initial Hamiltonian and a final Hamiltonian, alternating between the two.
This can be realized by restricting  the schedule function of VSQS to the form of bang--bang control and the cost functions to those of discrete optimization problems.
In VSQS, we group terms in the initial and final Hamiltonians and give each of them an independent schedule function.
Furthermore, we demonstrate VSQS using a navigator Hamiltonian in an example discrete optimization problem.
There are some methods in continuous optimization problems such as molecular systems which take similar form to QAOA
in the sense that an initial Hamiltonian and a final Hamiltonian
are applied in an alternation fashion. One example is Hamiltonian variational approach \cite{2015PhRvA..92d2303W}.
Molecular problems and discrete optimization problems are different in the sense that terms in the molecular Hamiltonians do not commute each other while
all the terms in Ising Hamiltonians commute each other.
As described in the following sections, grouping terms of molecular Hamiltonians 
and assigning them independent schedule functions provides the ability to generate accurate results within a short annealing time while keeping the number of variational parameters small.
The variational method we present in this paper incorporates these ideas to solve more general problems within a shortened annealing time per individual 
iteration.
We investigate the efficiency of the method in two 
contexts, the first being a quantum chemistry problem and the second an optimization problem. To clarify the main differences between the work investigated herein and the previous work done in the original VanQver paper~\cite{VanQver}, we reiterate that
the key point in that work was the introduction of an additional “navigator Hamiltonian” term which possesses a variationally determined coefficient. While the use of a navigator Hamiltonian still persists  in VSQS, we have chosen to investigate the advantages offered by instead determining the entirety of the scheduling functions for the remaining terms in the Hamiltonians,
$\hini$ and $\hfin$, variationally as well, not just that of the navigator Hamiltonian.

The structure of the paper is as follows.
In Sec.~\ref{Sec:VSQE}, we explain the variationally scheduled quantum simulation algorithm (\vrs). 
In Sec.~\ref{Sec:molecular}, we demonstrate the efficiency of \vrs~in solving the eigenstates problem for the hydrogen and P4 molecular systems.
In Sec.~\ref{Sec: Ising}, we apply \vrs~in solving the Ising model. 
In Sec.~\ref{Sec: Control errors}, we study how control error is mitigated in \vrs~by  
implementing an inaccurate final Hamiltonian into the algorithm's parameters. 
We conclude our work with a summary of our analysis in Sec.~\ref{Sec. Conclusion}.

\section{The Variationally Scheduled Quantum Simulation Algorithm}
\label{Sec:VSQE}

In the case of standard ASP, the time-dependent Hamiltonian $H$ has both a fixed initial Hamiltonian $\hini$ and final Hamiltonian  $\hfin$,
as well as their predetermined schedule functions $A(t)$ and $B(t)$, respectively:
\bea
H=A(t)\hini+B(t)\hfin
\label{standard ASP Hamil}
\eea
For a given annealing time $T$,
the coefficients $A(t)$ and $B(t)$ satisfy the following boundary conditions: \mbox{$A(0)=B(T)=1$} and $A(T)=B(0)=0$.
In the case of quantum annealing techniques that target Ising models, the Hamiltonian $\hini$ is
usually taken to be a summation of the transverse field for all the qubits: $\hini=\sum_{i}\sigma^{x}_i$.
In recent years, the efficiency of  quantum annealing for more-general functions has been investigated.
One approach is to increase the flexibility of the functions $A(t)$ and $B(t)$ themselves.
Instead of monotonically and smoothly changing functions, non-monotonic or non-smooth functions are considered \cite{Perdomo-Ortiz:2011,Chancellor:2016ys,DwaveReversAnnealTopological,ReverseAnnealLowRank,ReverseAnnealPortfolio,ReverseAnnealingONL,DynamicsofReverseAnneal,PowerofPausing,Pause.p-spin}.

The approach that we implement in \vtd~ is to find the optimal scheduling functions \textit{variationally} by using a quantum--classical hybrid method.
Let us first consider the case where $\hini$ and $\hfin$ both have a single coefficient each, $A(t,\bf{a})$ and $B(t,\bf{b})$:
\bea
H(t,\bf{a},\bf{b})=A(t,\bf{a})\hini +  B(t,\bf{b}) \hfin
\label{VSQS Hamil}
\eea
The schedule functions $A$ and $B$ are defined using variational parameters $\bf{a}=(a_1,a_2,\ldots)$ and $\bf{b}=(b_1,b_2,\ldots)$.
As one example of defining the functions by using variational parameters, 
we split the annealing time $T$ into $S$ intervals, $(i-1){T\over S}<t \le i{T\over S}$, $i\in[1,S]$.
At the end of the $i$-th interval, the schedule functions $A$ and $B$ take parameters $a_i$ and $b_{i}$, after which they are linearly interpolated 
in the intervals:
\bea
&&A(t,\bf{a})={a_{i}-a_{i-1}\over T/S} \left(t-(i-1){T\over S} \right)+a_{i-1} \cr
&&B(t,\bf{b})={b_{i}-b_{i-1}\over T/S} \left(t-(i-1){T\over S} \right)+b_{i-1}
\eea
In the above equations, the parameters $a_0=b_S=1$ and $a_S=b_0=0$ in order to satisfy the required boundary conditions.
Therefore, we do not treat them as variational parameters.
Another example of the schedule function has only two values, 0 and 1,
where the duration of each operation is determined variationally.
This so-called ``bang-bang'' control procedure is known to be optimal in the case of classical systems, according to Pontryagin's principle. However, the bang-bang control technique may or may not be optimal for quantum systems under certain conditions~\cite{QPontryagin,Qpontryagin2}.

\begin{figure*}[ht]
\centering
\includegraphics[scale=0.40]{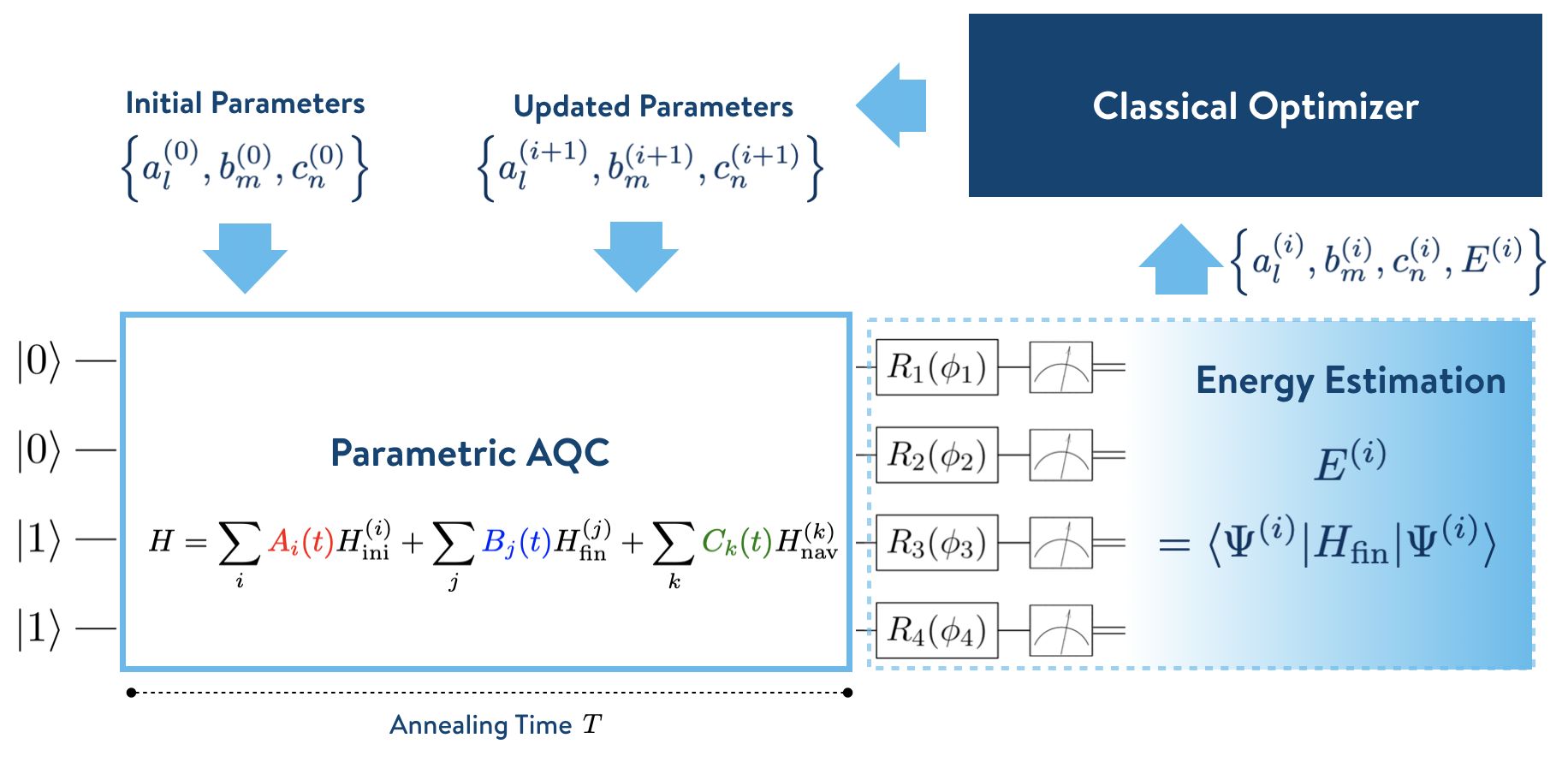}
\caption{
Visual representation of the structure of VSQS. First, generate an initial set of the variational parameters $(\bf{a}^{(0)}, \bf{b}^{(0)}, \bf{c}^{(0)})$, then 
 perform the parametric AQC $H(t,\bf{a}^{(0)}, \bf{b}^{(0)},\bf{c}^{(0)})$. From the generated quantum state, the expectation value of the final Hamiltonian, $E=\langle \hfin \rangle$, is obtained via the execution of measurements. Lastly, a classical optimizer is used in order to update the variational parameters of the system and reiterate the algorithm until convergence of the final Hamiltonian energy has been achieved.
}
\label{fig:InhomoVanQ loop}
\end{figure*}

The structure of \vtd~is shown in Fig.~\ref{fig:InhomoVanQ loop}.
To run VSQS, we first generate an initial set of the variational parameters $(\bf{a}^{(0)}, \bf{b}^{(0)})$, and 
subsequently perform the parametric ASP
$H(t,\bf{a}^{(0)}, \bf{b}^{(0)})$,
\bea
&&|\psi(T,\bf{a}^{(0)}, \bf{b}^{(0)})\rangle
=\mathcal{T}\exp\left(-i\int_{0}^{T} H(t,\bf{a}^{(0)}, \bf{b}^{(0)}) dt \right)|\psi(0)\rangle, \cr
&&
\label{eq: AQC quantum state}
\eea
where $\mathcal{T}$ represents the time ordering operator.
From the generated quantum state, the expectation value of the
final Hamiltonian, $E=\langle \hfin \rangle$, is obtained via the execution of measurements.
In the case of combinatorial optimization problems, $\hfin$ is a function of  $\sigma^{z}$ only; therefore, one measurement is sufficient for an adequate evaluation of the energy.
On the other hand, in the case of quantum problems, terms in $\hfin$ do not necessarily commute with each other, and the ground state of $\hfin$ is therefore not necessarily an eigenstate of each individual term. As a consequence, we need to measure the expectation value of each operator by repeatedly preparing the state.
It is worth mentioning that terms that qubit-wise commute with each other can be evaluated at the same time. 
Notably, various devices  allow measurements to be performed only in the computational basis. 
Thus, in order to measure terms containing $\sigma^{x}$ or $\sigma^{y}$, single-qubit rotations for the corresponding qubits need to be performed at the end of the annealing process so that 
the measurement of the desired term becomes a tensor product of $\sigma^{z}$ and the identity operator~$I$.
Recall that the expectation value $\langle \hfin \rangle$ is a function of the variational parameters, and 
therefore optimal variational parameter values must be chosen if $\langle \hfin \rangle$ is to yield a meaningful result.
The data representing the obtained energy and the variational parameters are sent as input to a classical optimizer,  
which then updates the values $(\bf{a}^{(1)}, \bf{b}^{(1)})$.
Quantum annealing is performed based on the updated values $(\bf{a}^{(1)}, \bf{b}^{(1)})$, and the calculations are iterated until the convergence condition of the energy has been satisfied.

\begin{figure}[ht]
\centering
\includegraphics[scale=0.4]{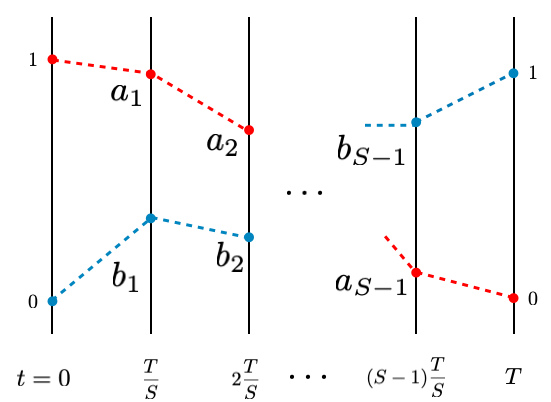}
\caption{
Schematic of the variationally determined time dependence employed in VSQS, depicting the variational parameters $\bf{a}=(a_1,a_2,\ldots)$ and $\bf{b}=(b_1,b_2,\ldots)$ that are used in order to define the variationally determined schedule functions $A$ and $B$. The annealing time $T$ is split into $S$ intervals, $(i-1){T\over S}<t \le i{T\over S}$, $i\in[1,S]$, which is one example of defining the functions by using variational parameters.
}
\label{fig:VTD}
\end{figure}

Another example implementation of  \vrs~we consider is one that  provides an independent schedule function for each term. In employing quantum annealing to target Ising problems, in contrast to using a single coefficient $A(t)$ 
for all of the transverse fields, we instead consider
the qubit-dependent coefficient $A_{i}(t)$:
\bea
H=\sum_{i} A_{i}(t)\sigma^{x}_{i} +B(t)\hfin
\eea
This type of protocol is commonly denoted ``inhomogeneous transverse field scheduling'', or ``annealing offset scheduling''.
Theoretical research has 
suggested that dramatic improvements in the likelihood of success can be accomplished by applying qubit-independent transverse fields~\cite{InhomogeneousMcMahon,Inhomo-DicksonAmin,InhomoDriver,Inhomo-pspin}.
To extend this idea toward solving more-general problems, one can employ term-dependent scheduling in \vrs.
To achieve this, let us write
\bea
&&\hini=\sum_{i=1}^{M_{\text{ini}}} J_{\text{ini},i} \bf{\sigma}_{\text{ini}}^{i},~
\hfin=\sum_{j=1}^{M_{\text{fin}}} J_{\text{fin},j} \bf{\sigma}_{\text{fin}}^{j},~
\eea
where  ${M_{\text{ini}}}$ and ${M_{\text{fin}}}$ are the numbers of terms in $\hini$ and $\hfin$, respectively, 
$\bf{\sigma}^{i}$ is a tensor product of Pauli matrices,
and $J_{\text{ini},i}$, $J_{\text{fin},j}$ are coefficients.
In \vtd,
the time dependence of each term is defined by the variational parameters $\bf{a}_{i},\bf{b}_{j}$:
\bea
H=\sum_{i=1}^{M_{\text{ini}}} A_{i}(t,\bf{a}_{i}) J_{\text{ini},i} \bf{\sigma}_{\text{ini}}^{i}
+\sum_{j=1}^{M_{\text{fin}}} B_{j}(t,\bf{b}_{j}) J_{\text{fin},j} \bf{\sigma}_{\text{fin}}^{j}
\eea
Note that some of the coefficients can be the same function, for example, $\bf{a}_{i}=\bf{a}_{j}$ for some $i\neq j$.
This reduces the number of variational parameters and therefore reduces the computational cost of the classical optimizer.
In what follows, we consider distributing $\bf{\sigma}_{\text{ini}}^{i}$ and $\bf{\sigma}_{\text{fin}}^{j}$ into $I$ and $F$ groups ($I$ and $F$ are integers), respectively, and give each group 
an independent schedule function.\\
\\Finally, we consider the addition of terms to \vrs. While it is standard to choose the final Hamiltonian as the problem Hamiltonian in whose ground state we are interested, there is no restriction in what 
kind of terms are switched on during annealing.
In \cite{VanQver,Farhi:differentPaths,Perdomo-Ortiz:2011,crosson2014different,VQE3}, the additional terms were introduced in order to improve computational performance.
We call the set of additional terms a navigator Hamiltonian $\hnav$.
The only condition that must be satisfied is that the schedule function of $\hnav$ is zero at both the beginning and the end of the annealing process,
\bea
&H(t,\bf{a},\bf{b},\bf{c})=A(t,\bf{a})\hini +  B(t,\bf{b}) \hfin + C(t,\bf{c}) \hnav\,, \cr
&
\eea
with $C$ satisfying the boundary conditions \mbox{$C(0)=C(T)=0$}, and a nontrivial choice of $\hnav$.
In \cite{VanQver}, $\hnav$ was  chosen to be a cluster operator that was used  in either the
unitary coupled-cluster (UCC) or generalized unitary coupled-cluster (GUCC) method. 
In example combinatorial optimization problems, non-standard quantum fluctuations such as $\sigma^{x}_i\sigma^{x}_j$ interactions were used as navigator Hamiltonians \cite{crosson2014different,Seki2012,Seoane2012,Seki:2015,Nishimori:2016aa,PhysRevB.95.184416,2017PhRvA95d2321S,PhysRevA.99.032315,2019PhRvA99d2334A}.


\section{Application to molecular systems}
\label{Sec:molecular}

\subsection{Hydrogen Molecule}
We demonstrate the efficiency of \vrs~ within the context of determining the ground states of various molecular systems via a direct comparison with results obtained using the standard ASP approach. 
We use QuTiP \cite{qutip1,qutip2} for solving the Sch{\"o}dinger equation and the Lindblad master equation.
In the following examples, the number of variational parameters used in the algorithm are determined by the split number of the annealing time ($S$), the group number of the initial Hamiltonian $(I)$, and the group number of the final Hamiltonian $(F)$.
As an initial example, we consider a hydrogen molecule, whose Hamiltonian 
takes the form
\bea
\hfin=f_0+f_1(\sigma^{z}_1+\sigma^{z}_2)+f_3 \sigma^{z}_1\sigma^{z}_2
+f_4 \sigma^{x}_1\sigma^{x}_2\,,
\eea
where we use the Bravyi--Kitaev transformation~\cite{Bravyi:00} of the second quantized Hamiltonian
and
remove two qubits based on the conservation of the spin \mbox{symmetries~\cite{TaperingOffQubits,GoogleHydrogen}.}
The coefficients $f_i$ are functions of the nuclear separation distance $d$.
As this is a two-qubit problem with only $\sigma_i^{z}\sigma_j^{z}$ and $\sigma_i^{x}\sigma_j^{x}$  two-qubit couplings, 
it is feasible for this model to be implemented on a hardware platform such as~\cite{AdiabaticIBM}.
More general chemical problems usually possess higher-order couplings, and in such cases 
the use of perturbative gadgets to reduce these higher-order couplings to two-qubit couplings may be required. For more detail on such cases, see~\cite{kempe:1070,Jordan:08,AQC-for-qchem,Bravyi-KitaevSeeley,Tranter-BravyiKitaev}.
The initial Hamiltonian for the system is taken as the Hartree--Fock Hamiltonian
\bea
\hini=g_1(\sigma^{z}_1+\sigma^{z}_2).
\eea
Again, the coefficients $g_i$ are functions of the nuclear separation distance $d$.
In order to see the roles played by time splitting and grouping terms 
in~\vrs, we study three different cases of 
the split number, the initial group number, and the final group number:
$(S,I,F)=(2,1,1),~ (5,1,1)$, and $(5,2,4)$.
In the case of $(S,I,F)=(2,1,1)$ or $(5,1,1)$, we do not split the terms in the initial and the final Hamiltonians. In this sense, they are homogeneous. The split number $S=2$ is the simplest nontrivial schedule and $S=5$ has more flexibility.
In the case of $(S,I,F)=(5,2,4)$, the number of $I$ and $F$ are chosen to equal the maximum: the number of terms in the initial Hamiltonian is 2 and that of the final Hamiltonian, except the term proportional to the identity, is 4.  
In this sense, it is maximally inhomogeneous.
For $(S,I,F)=(2,1,1)$ and $(5,1,1)$, the time-dependent Hamiltonian is given by
\bea
H=A(a_i)\hini+B(b_i)\hfin\,,
\eea
with $i=1$ and $i \in [1,4]$ for  $(2,1,1)$ and $(5,1,1)$,
respectively.
For $(5,2,4)$, all terms have the unique time dependence
\bea
H=\sum_{k=1}^{2}A_k(a_i)\sigma^{z}_k+
\sum_{j=1}^{4}B_j(b_i)\bf{\sigma}_{j}\,,
\eea
where $\bf{\sigma}_{j}=\{\sigma^{z}_{1},\sigma^{z}_{2},\sigma^{z}_{1}\sigma^{z}_{2},
\sigma^{x}_{1}\sigma^{x}_{2}\}$.

The nuclear separation distance 
is chosen to be \mbox{$d=1~\ang$}, and 
the  amplitudes $|A_{k}|$ and $|B_{k}|$ are bounded by the value $10$.
To set the range for the variational parameters, we use the
limited-memory \mbox{Broyden--Fletcher--Goldfarb--Shanno} method for bound-constrained optimization (L-BFGS-B) as an optimizer.
The main purpose of this constraint in the amplitude of the variational parameters is to avoid a large deviation in the norm of the Hamiltonian from that of the standard ASP case.
As a comparison, standard ASP is considered with the schedule functions
\bea
A(t)=1-\left({t\over T}\right)^2,~~B(t)=\left({t\over T}\right)^2.
\eea

\begin{figure}[ht]
\centering
\includegraphics[scale=0.35]{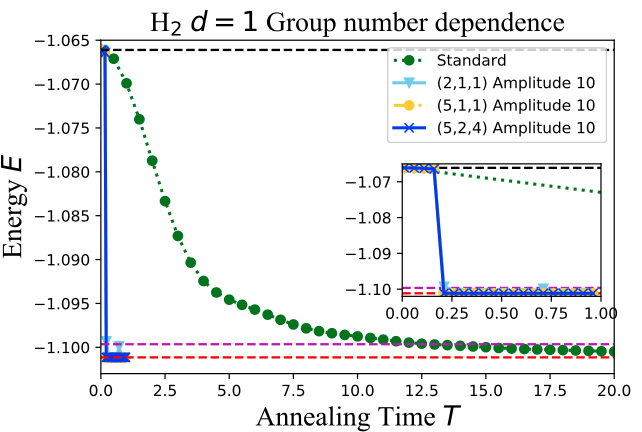}
\caption{
Obtained energy $E$ as a function of the annealing time $T$. The dotted green line shows the result obtained from standard ASP, and the solid light-blue, dotted yellow, and solid dark-blue lines are the results of VSQS with $(S,I,F)=(2,1,1),~(5,1,1)$, and $(5,2,4)$, respectively. The amplitudes $|A_k|$ and $|B_k|$ are bounded by 10.
The yellow line for $(5,1,1)$ overlaps almost exactly with the blue line for $(5,2,4)$. 
}
\label{fig:EvsT_inhomo_standard_H2}
\end{figure}

The obtained energy as a function of the annealing time $T$ is shown in Fig.~\ref{fig:EvsT_inhomo_standard_H2}. The tolerance of the optimizer is set to $10^{-6}$. 
The optimizer converges between $50$--$75$ iterations for $T<0.5$
and $25$--$50$ iterations for $T>0.5$.
As a metric for the accuracy of the obtained results, we use the chemical accuracy (1~kcal/mol).
The adiabatic theorem guarantees that the obtained values will converge to the exact result for a given system, in the case when the annealing time $T$ is taken to be large, and in the absence of noise and errors.
We consider the required annealing time to achieve chemical accuracy $T_{\text{CA}}$ as a measure of efficiency.
We emphasize that this is the annealing time per individual run, since this is the relevant quantity for obtaining accurate results on noisy quantum devices, not the total annealing time.

The numerical results show that
in the example of a hydrogen molecule, there is no clear  difference in accuracy between different groupings. In all three cases studied, the value of the energy drops rapidly from
the Hartree--Fock energy to the exact energy at $T_{\text{CA}}\simeq 0.2$.
On the other hand, it decreases continuously in the case of  standard ASP
and achieves  chemical accuracy at $T_{\text{CA}}\simeq 12.9$.

\begin{figure}[ht]
\centering
\includegraphics[scale=0.35]{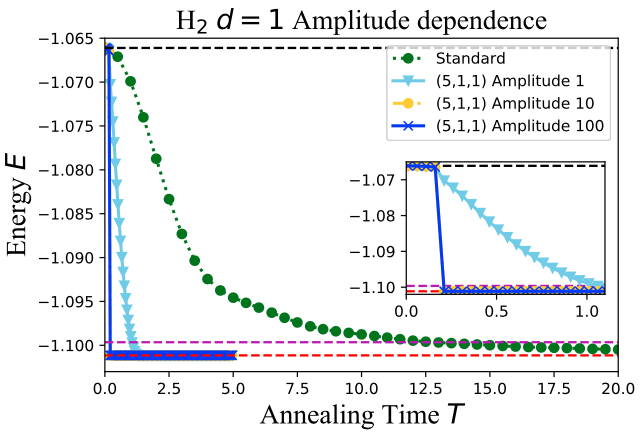}
\caption{
Obtained energy $E$ as a function of the annealing time $T$.
The dotted green line is the result obtained from standard ASP, and the solid light-blue, dotted yellow, and solid-blue lines are
the results of VSQS with
$(S,I,F)=(5,1,1)$ and the amplitudes bounded by 1, 10, and 100, respectively.
The yellow line for the amplitude 10 overlaps almost exactly with the blue line for the amplitude 100.
}
\label{fig:EvsT_H2_A-dependence}
\end{figure}

Next, we investigate the dependence of the accuracy on the amplitude.
For a fixed group number $(5,1,1)$, we change the upper bound of the amplitudes.
The main objective is to understand whether the achievement of the shorter annealing time is due to the large amplitude of the Hamiltonian.
We study three upper bounds: 1, 10, and 100.
The obtained energy as a function of the annealing time is shown in Fig.~\ref{fig:EvsT_H2_A-dependence}.
For the case of the amplitude with a bound of 1, the energy decreases smoothly from the Hartree--Fock energy to the exact energy with a time to chemical accuracy of $T_{\text{CA}}\simeq 1.1$.
On the other hand, the obtained energies for the cases of amplitudes bounded by 10 and 100 are the same, and they suddenly drop from the Hartree--Fock energy to the exact energy at $T_{\text{CA}}\simeq 0.2$. This shows that the optimal schedule functions are
within the range of $\pm 10$ for the entire $t\in [0,T]$ and an increase in the 
bound does not improve the performance.
Notice that since we set boundary conditions for the amplitudes at $t=0$ and $T$,
for a given set of group numbers, one cannot simply rescale the Hamiltonian to
shorten the annealing time.

\begin{figure}[ht]
\centering
\includegraphics[scale=0.35]{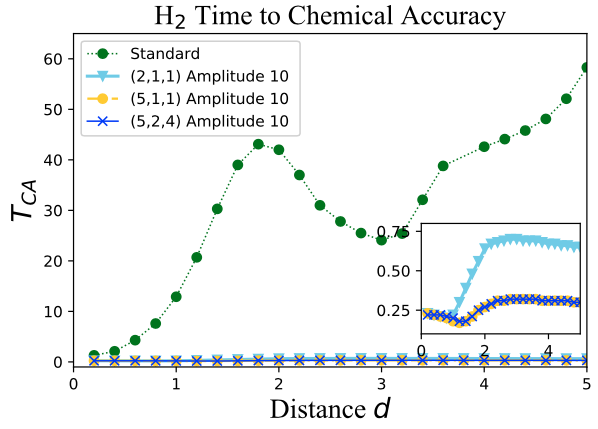}
\caption{Time to chemical accuracy for H$_\text{2}$.
The dotted green line is the result of standard ASP, while the solid light-blue, dotted yellow, and solid dark-blue lines are the results of VSQS with
$(S,I,F)=(2,1,1),~(5,1,1)$, and $(5,2,4)$, respectively.
}
\label{fig:TCA_H2}
\end{figure}

The required annealing time to achieve chemical accuracy changes as the nuclear separation distance 
changes. 
The $T_{\text{CA}}$ for various nuclear separation distances is shown in Fig.~\ref{fig:TCA_H2}.
For standard ASP, $T_{\text{CA}}$ tends to increase as the nuclear separation distance increases. The same feature is observed in~\cite{AdiabaticIBM}.
$T_{\text{CA}}$ decreases between $d=1.8$~$\ang$ and $3$~$\ang$.
As explained in the Appendix, the obtained energies show some oscillation as a function
of the annealing time $T$. This suggests that the annealing process uses non-adiabatic transitions to 
reach accurate results in a shorter time than would otherwise be expected from the adiabatic condition. See, for instance,~\cite{PhysRevA.88.013818,Messina_2014,
PhysRevA.95.032335,2019arXiv190300574M}. 
In the case of \vrs, $T_{\text{CA}}$ has a different dependence on $d$ compared to the
standard ASP case. It takes the maximum values between $d=2.8$ and $3.0$ for (2,1,1)
with $T_{\text{CA}}\simeq 0.7$, whereas they are between $d=2.8$ and $3.6$ for (5,1,1) and
(5,2,4) with $T_{\text{CA}}\simeq0.32$. Comparatively, the  $T_{\text{CA}}$ is small for standard ASP at these distances due to the non-adiabatic transitions.

\begin{figure}[ht]
\centering
\includegraphics[scale=0.35]{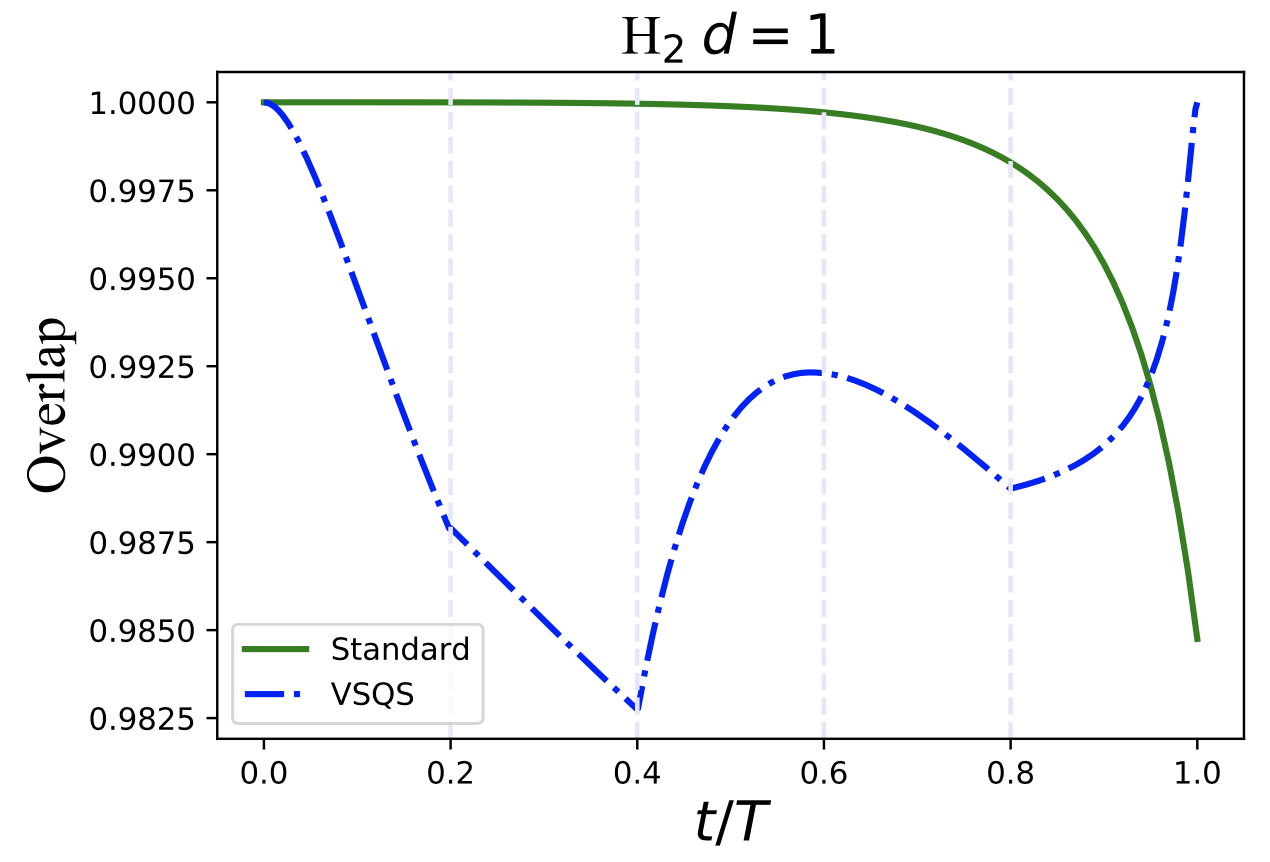}
\caption{Overlap between the wavefunction generated
by the time-dependent Hamiltonian and the instantaneous ground state for $\text{H}_2,~d=1.0~\ang,$ and $T=0.25$.
For VSQS, the relevant parameters are set to be $(S,I,F)=(5,1,1)$.
The green line represents the standard ASP result and the blue dash-dotted line represents the VSQS result.
}
\label{fig:ovrelap}
\end{figure}

In order to understand the mechanism for shortening the annealing time, we study how close the quantum annealing follows the adiabatic evolution.
The quantum state generated in VSQS or in standard ASP at time $t$ is given by
\bea
|\psi(t)\rangle = \mathcal{T}\exp
\left(
-i\int_{0}^{t}H(s)ds
\right)|\psi(0)\rangle,
\eea
where $H(t)$ is (\ref{standard ASP Hamil}) for standard ASP and (\ref{VSQS Hamil}) for VSQS.
The instantaneous ground state $|GS(t)\rangle$ is given by the lowest-energy state of the Hamiltonian at time $t$,
\bea
H(t)|GS(t)\rangle = E_0(t)|GS(t)\rangle.
\eea
Again, $H(t)$ is chosen to be (\ref{standard ASP Hamil})
or (\ref{VSQS Hamil}) for  standard ASP or VSQS, respectively.
The overlap
$\big| \langle GS(t) | \psi(t)\rangle \big|$ determines how closely the 
evolution follows the adiabatic evolution.
In Fig.~\ref{fig:ovrelap}, we show the overlap for $\text{H}_2$, $d=1.0$~$\ang$, and $T=0.25$. For  
VSQS, the split numbers are chosen as $(S,I,F)=(5,1,1)$.
Standard ASP follows the adiabatic path in the beginning of the annealing; however, it does deviate away from this towards the end of the calculation. On the other hand, VSQS follows the diabatic path, and some portion of the wavefunction shifts to an excited state. Subsequently, the whole wavefunction comes back to the ground state towards the end of the annealing process. In other words, VSQS partially accesses excited states in order to realize its observed speedup.
A similar feature was observed during the investigation of the VanQver algorithm \cite{VanQver}. 

\subsection{P4 Molecule}

\begin{figure}[ht]
\centering
\includegraphics[scale=0.35]{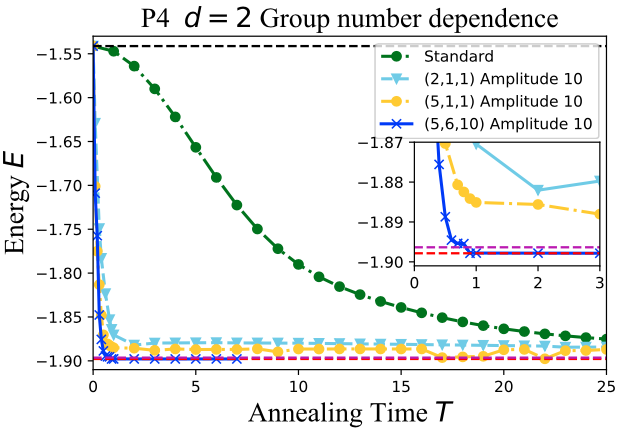}
\caption{
Obtained energy $E$ as a function of the annealing time $T$ for the P4 molecule with a distance $d=2~\ang$.
The dotted green line represents the result obtained from standard ASP, and the solid light-blue, dotted yellow, and solid dark-blue lines show the results of VSQS with
$(S,I,F)=(2,1,1),~(5,1,1)$, and $(5,6,10)$, respectively.
The amplitudes $|A_k|$ and $|B_k|$ are bounded by 10.
}
\label{fig:P4_EvT_D2}
\end{figure}

The second example system we investigate is P4, a system of two hydrogen molecules with bonds lying parallel to each other.
In particular, we choose a square configuration where each hydrogen atom is located at a vertex and the separation between each edge is $2$~$\ang$. 
In symmetry-conserving Bravyi--Kitaev transformation, 
the Hilbert space dimension of this molecule is $2^6$.
One motivation for considering this system is that
it is difficult to obtain an accurate energy value with this configuration because of  degeneracy.
It was shown in~\cite{VanQver} that the  classical  method with coupled-cluster  singles  and  doubles (CCSD), as well as the variational quantum eigensolver~(VQE) with unitary coupled-cluster singles and doubles (UCCSD) both fail to achieve chemical accuracy.
We study the performance of~\vrs~for the fixed amplitude bounds $|A_i|,~|B_i|\le10$.
Three choices for the group numbers $(S,I,F)=(2,1,1),~(5,1,1)$, and $(5,6,10)$
are considered.
The ~results are shown in Fig.~\ref{fig:P4_EvT_D2}.
In  standard ASP (shown using a dotted green line), the energy decreases monotonically as a function of the annealing time, and requires $T \ge 456$ to achieve chemical accuracy.
For all the cases using the variational method, the energy decreases much faster than with standard ASP, yet the details of the decrease in energy are different from the case of hydrogen.
For the hydrogen molecule, there was no discernible difference despite the choice of group number.
However, the difference is clear in the case of P4, where the group number $(5,1,1)$ provides a more stable configuration compared to that of $(2,1,1)$. Therefore, it is readily apparent that the increase of the 
split number is non-trivial. Moreover, the result for the group number 
$(5,6,10)$ reaches chemical accuracy around 
$T_{\text{CA}}\simeq 0.9$, whereas those of $(2,1,1)$ and $(5,1,1)$ plateau for a certain range of $T$ before achieving chemical accuracy around
$T_{\text{CA}}\simeq 133$
and
$T_{\text{CA}}\simeq 23$, respectively.

In general, it is challenging to estimate the number of required parameters to achieve a required accuracy.
In the case of the unitary coupled-cluster singles and doubles (UCCSD) ansatz in VQE, the number of parameters scales as
$\mathcal{O}(n^2(\eta -n)^2)$ where $n$ is the number of electrons and $\eta$ is the number of total spin-orbitals.
However, the achievable accuracy within UCCSD is system dependent.
Employing algorithms such as ADAPT VQE~\cite{2019NatCo..10.3007G,2019arXiv191110205L}
or qubit coupled cluster (QCC)~\cite{2018arXiv180903827R,2019arXiv190611192R}
keeps improving the computational accuracy by adding entanglers.
Since the ansatz is not fixed in these methods, it is possible to achieve chemical accuracy.
However, the relation between the number of entanglers 
and the accuracy of obtained results has been studied only numerically.
The uncertainty of the scaling of parameter numbers in VSQS is 
similar to that of the ADAPT VQE and QCC algorithms.
It is challenging to explore the scaling of parameters
on classical computers since it becomes it quickly becomes intractable to simulate a time-dependent Schr\"{o}dinger equation as the system size increases.

\section{Ising model}
\label{Sec: Ising}

\begin{figure}[ht]
\centering
\includegraphics[scale=0.3]{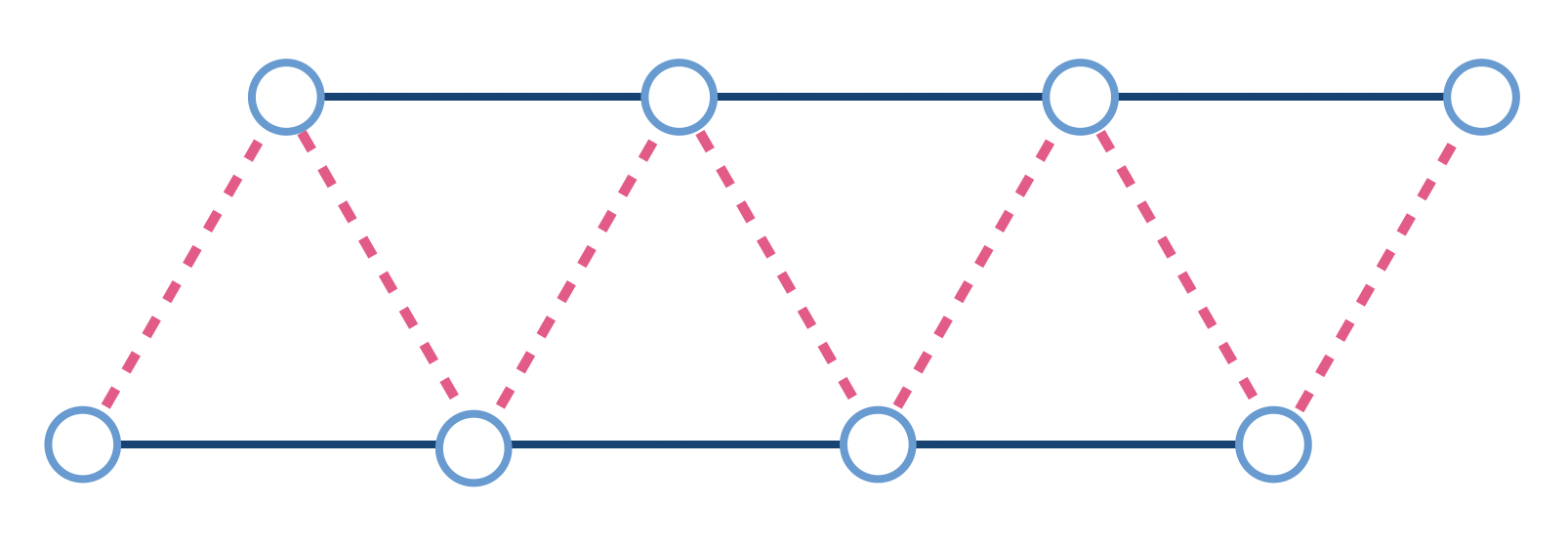}
\caption{Triangular lattice. Circles represent qubits, and the solid dark-blue and the dashed red lines represent antiferromagnetic and ferromagnetic couplings, respectively. 
}
\label{fig:TriangleLattice}
\end{figure}

In this section, we apply \vrs~for solving an Ising model problem.
As an example, we consider an eight-qubit triangular lattice, shown in
Fig.~\ref{fig:TriangleLattice}.
Each vertex (shown using a light-blue circle) represents a qubit, 
each solid dark-blue line (within the upper layer or lower layers) represents antiferromagnetic coupling ($J^{AF}>0$),
and each edge of dashed red line (inter-layer) represents a ferromagnetic coupling ($J^{F}<0$).
The values of the ferromagnetic and antiferromagnetic
couplings are randomly generated:

\bea
\hfin=\sum_{\mathclap{(i,j) \in \text{Intralayer}}}\; J^{AF}_{ij}\sigma^{z}_i\sigma^{z}_j
+\sum_{\mathclap{(k,l) \in \text{Interlayer}}}\; J^{F}_{kl} \sigma^{z}_k\sigma^{z}_l 
\eea
\noindent
where $\{i,j,k,l\}$ take values in $\{1,2,\ldots, 8\}$.
We consider the following Hamiltonian: 

\bea
H=A(t)\hini +B(t)\hfin +\sum_{\mathclap{(i,j)\in \text{Edges}}}\; C_{ij}(t)\sigma^{x}_i \sigma^{x}_j
\eea

We take $M_{\text{ini}}=M_{\text{fin}}=1$ while all the terms in $\hnav$ have 
unique time dependence, and choose to investigate the split numbers 2 and 5.

\begin{figure}[ht]
\centering
\includegraphics[scale=0.4]{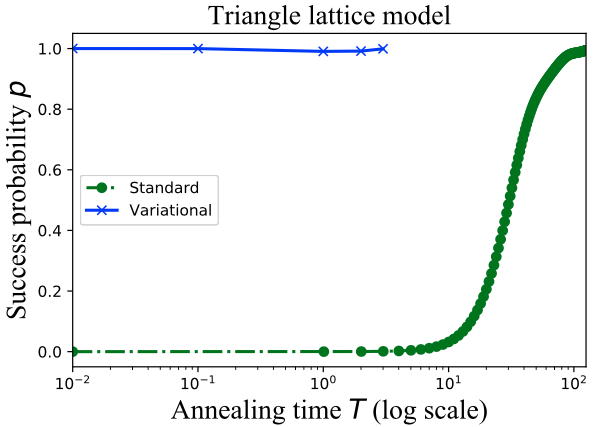}
\caption{Success probability $p$ of  standard ASP and \vrs~for the triangular lattice Ising model.
VSQS yields a high success probability even when the corresponding success probability with standard ASP is close to zero. 
}
\label{fig:SuccessvsT_Ising_N8_Standard}
\end{figure}

First, we look at the case where $S=5$. Figure~\ref{fig:SuccessvsT_Ising_N8_Standard} shows the success probability of
 standard ASP and \vrs~as a function of the annealing time $T$. In the standard ASP case, the success probability remains close to zero until $T\sim10$, then begins to increase.
At $T\simeq 116$, the success probability reaches $0.99$.
On the other hand, in~\vrs, the probability is greater than $0.99$ even when the annealing time is as short as $0.01$.

\begin{figure}[ht]
\centering
\includegraphics[scale=0.4]{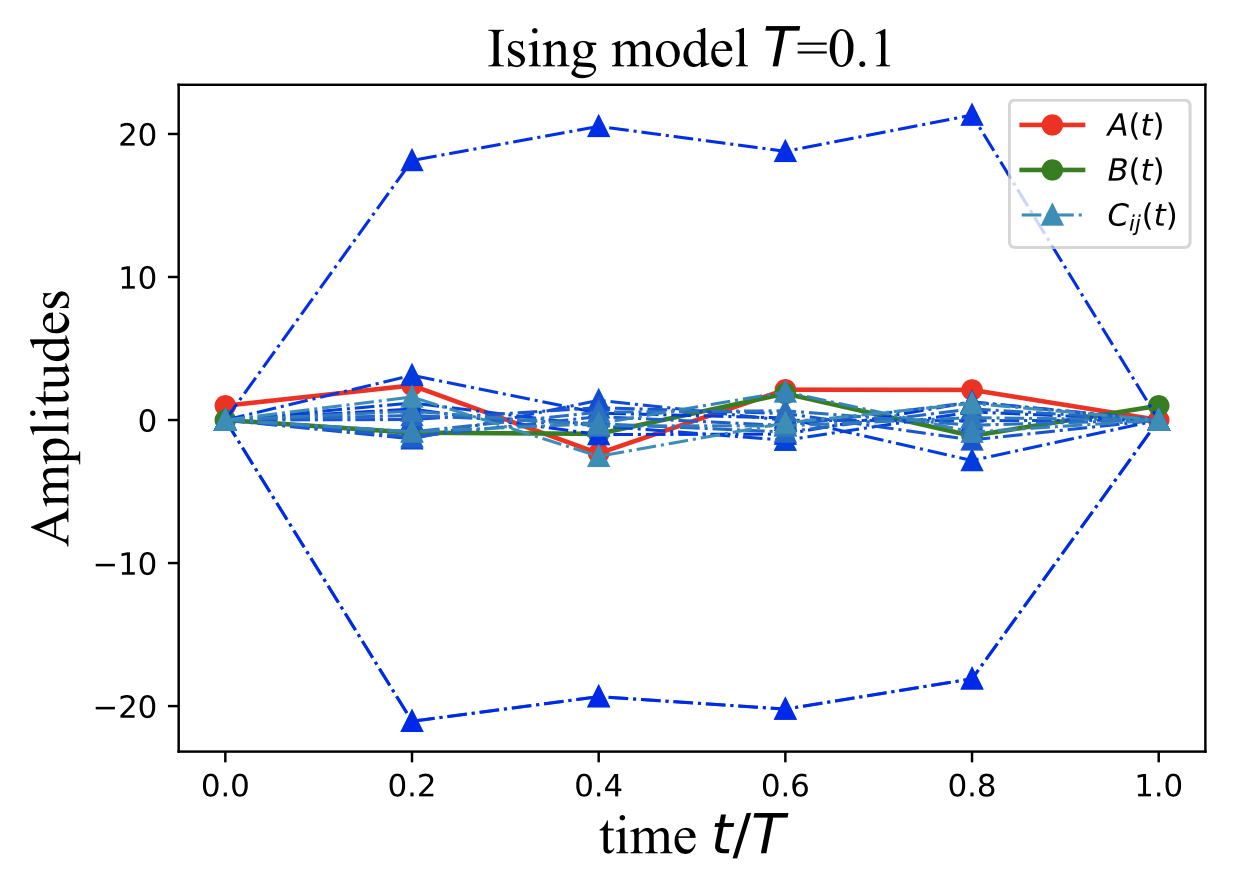}
\caption{Schedule functions of $T=0.1$ for the Ising model. The schedule functions of the navigator Hamiltonian take both positive and negative values. Therefore, the Hamiltonian is non-stoquastic.
}
\label{fig:Amp_IsingN8_T01-S5I1F1XX}
\end{figure}

\begin{figure}[ht]
\centering
\includegraphics[scale=0.4]{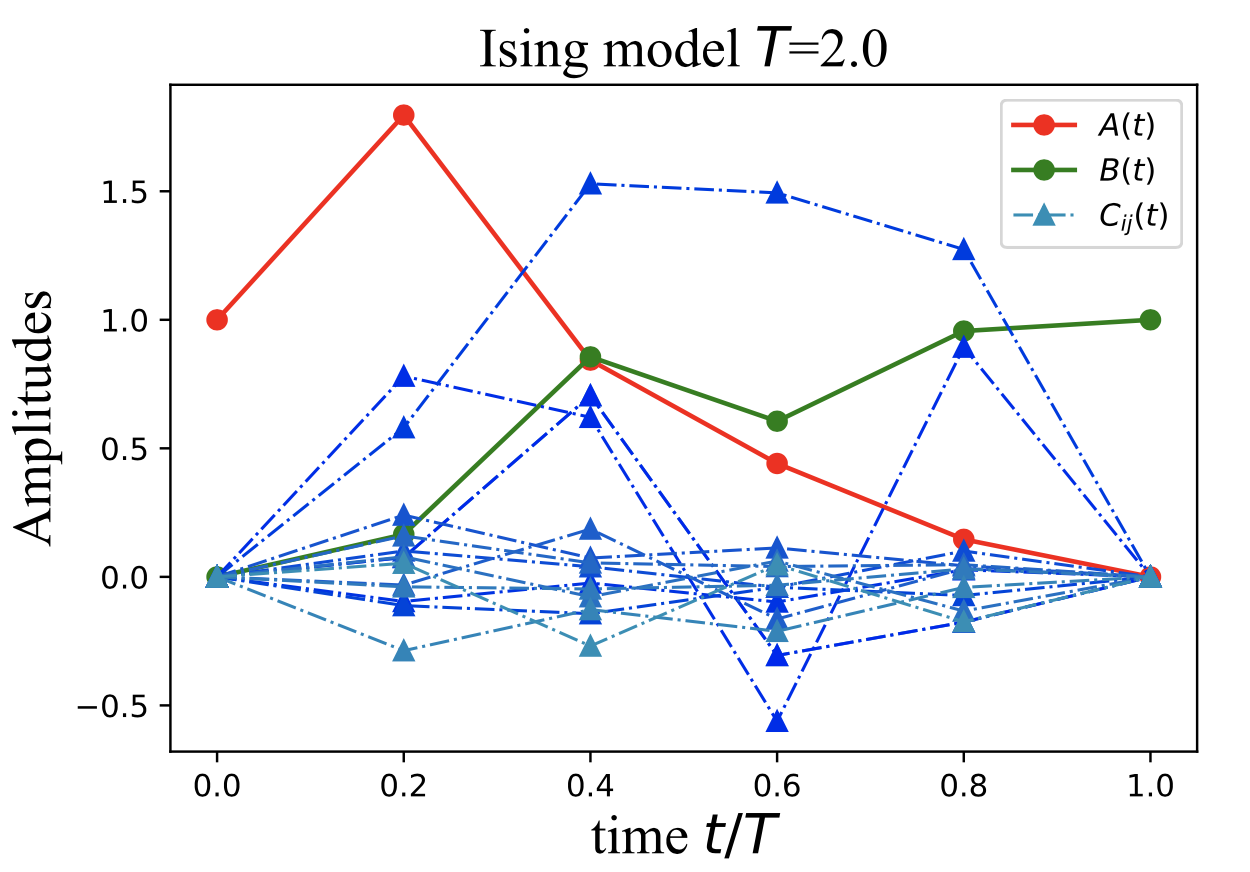}
\caption{Schedule functions of $T=2.0$ for the Ising model.
 Some schedule functions take both large positive and negative values.
Therefore, the Hamiltonian is non-stoquastic.
}
\label{fig:Amp_IsingN8_T2-S5I1F1XX}
\end{figure}

Figure~\ref{fig:Amp_IsingN8_T01-S5I1F1XX} and Fig.~\ref{fig:Amp_IsingN8_T2-S5I1F1XX} 
show the optimal schedule functions $\{A(t),B(t),C_{ij}(t)\}$ for specific annealing times $T=0.1$ and $T=2.0$, respectively.
The solid red line represents the coefficient $A(t)$ of $\hini$, and 
the  solid green line represents the coefficient $B(t)$ of $\hfin$.
The dashed light-blue  lines represent the coefficients $C_{ij}(t)$ of the terms in $\hnav$.
At $T=0.1$, two couplings in $\hnav$ become large during  annealing, while the others 
remain small.
As the annealing time increases, all the schedule functions take values within a small parameter region.
The schedule functions $C_{ij}(t)$ fluctuate between both positive and negative values. This means that the Hamiltonian is non-stoquastic (A Hamiltonian is called stoquastic when all off-diagonal elements in the computational basis are non-positive).

Our motivation for using the variational method is to obtain  accurate results despite the fact that quantum devices are noisy.
Therefore, up until this point we have focused on the annealing time per individual run. However, we also study 
the total annealing time for the entire calculation when applying the variational approach. Namely, we take into account the repetition of the run to
obtain the expectation values as well as the
iteration of the optimization process.
To do this, we first consider standard ASP. We denote the success probability at an annealing time $T_{\text{stand}}$ as $p_{\text{stand}}(T_{\text{stand}})$.
By repeating the same calculation $N_{\text{stand}}$ times, one can increase the $p_{\text{stand}}(T_{\text{stand}})$
using the expression
$1-(1-p_{\text{stand}}(T_{\text{stand}}))^{N_{\text{stand}}}$ with the total computational time being $T_{\text{stand}} N_{\text{stand}}$.
\begin{figure}[ht]
\centering
\includegraphics[scale=0.4]{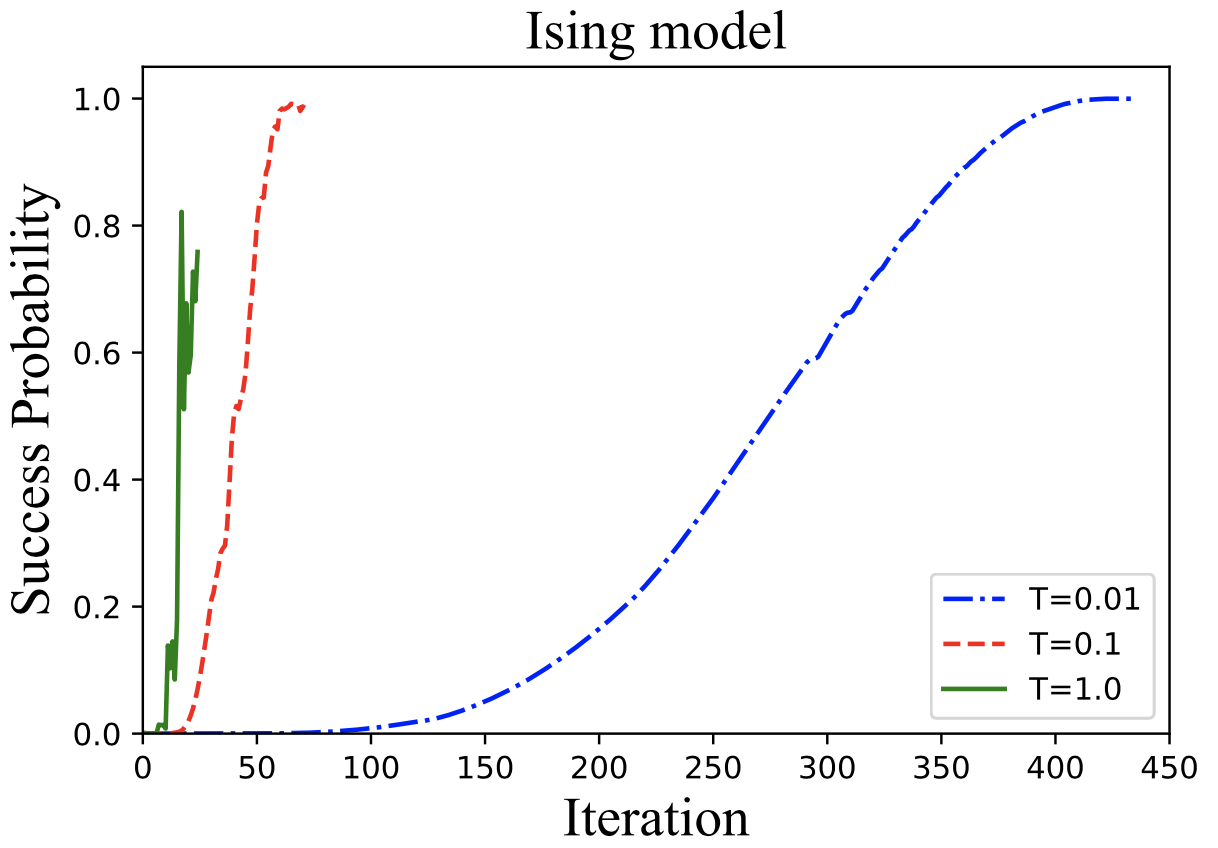}
\caption{Success probability as a function of iteration number. The split number $S = 2$. The initial values for $C_{ij}$ are chosen randomly from the interval $[-1,1]$.
}
\label{fig:OptimizationCobyla_Ising}
\end{figure}

In the case of \vrs, we must first iterate the performed calculations to optimize variational parameters.
Let us denote the number of iterations for optimizing the variational parameters by $N^{\text{opt}}_{\text{\vrs}}$, and the annealing time per individual run by $T_{\text{\vrs}}$. 
 In the case of variational methods, including a quantum approximate optimization algorithm (QAOA)~\cite{Farhi:2014ab}, which uses expectation values of the Hamiltonian to adjust variational parameters, one needs to repeat a given calculation in order to reduce statistical errors. We denote the
 repetition number for obtaining an expectation value of a Hamiltonian by $M$.
 Having $N^{\text{opt}}_{\text{\vrs}}$ iterations of the optimization process requires 
 $N^{\text{opt}}_{\text{\vrs}}M$ repetitions on a quantum device.
 The success probability of an individual run at $T_{\text{\vrs}}$ is denoted by $p_{\text{\vrs}}(T_{\text{\vrs}})$.
 Then, by repeating the calculation  $N^{\text{add}}_{\text{\vrs}}$ times with fixed variational parameters, we can 
 obtain the correct answer with a success probability of
$1-(1-p_{\text{\vrs}}(T_{\text{\vrs}}))^{N^{\text{add}}_{\text{\vrs}}}$ and with a total annealing time of
$T_{\text{vrs}}(N^{\text{opt}}_{\text{\vrs}}M+N^{\text{add}}_{\text{\vrs}})$.
The change of the success probability as a function of  iterations $N^{\text{opt}}_{\text{\vrs}}$ is given in Fig.~\ref{fig:OptimizationCobyla_Ising}.
We choose $S=2$ and consider random values taken from the interval $[-1,1]$ for the initial variational parameters $C_{ij}$.
The group number for the navigator Hamiltonian is 7.
One of the main differences between a quantum chemistry problem and a combinatorial optimization problem is that it is not necessary for the combinatorial optimization problem to achieve very high accuracy for a single run. 
So long as the success probability for a single run is reasonably high, one can increase the success probability quickly by repeating the calculations.
Therefore, we choose a low tolerance for the optimization: $tol=1.0$.
 We use the constrained optimization by linear approximation (COBYLA) algorithm. We see that the success probability increases within a relatively small number of iterations for annealing times $T_{\text{\vrs}}=0.1$ and $1.0$.
For instance, for $T_{\text{\vrs}}=1.0$, the optimization converges to 
$0.759$ after $24$ iterations; 
 for $T_{\text{\vrs}}=0.1$, the optimization converges to 
$0.985$ after $71$ iterations; and
 for $T_{\text{\vrs}}=0.01$, the optimization converges to 
$1.0$ after $434$ iterations.
Therefore, the total annealing time needed to find the correct result with a success probability of $99 \%$ is 
$4.34M$ for $T_{\text{\vrs}}=0.01$,  
$7.2M+0.1$ for  $T_{\text{\vrs}}=0.1$, and
$24M+4$ for $T_{\text{\vrs}}=1.0$ (i.e., $4$ iterations after the optimization converges). 
For reference, the total annealing time needed for standard ASP to achieve a success probability of $99\%$ is $T_{\text{stand}}=116$.

\section{Errors}

\subsection{Systematic Control Error}
\label{Sec: Control errors}

To date, many experiments (for instance, \cite{PeruzzoPhotonicPQ,GoogleHydrogen,Hardware-efficientIBM}) 
have demonstrated that the variational quantum eigensolver (VQE) is robust against systematic control errors.
In VQE, quantum circuits are characterized by variational parameters. 
When the quantum gates over-rotate or under-rotate qubits due to inaccurate control, a classical optimizer finds different input angles that lead to more-accurate qubit rotations on quantum devices.

In ASP, a critical control error may occur in the couplings of the final Hamiltonian $\hfin$.
In the ideal case, the quantum state reaches the ground state of $\hfin$ at the end
of the annealing process. However, without error correction, the couplings on a  quantum device
may be degraded by control errors. In this case, the final state will be a ground state of an \textit{inaccurate} Hamiltonian. Approaches such as using a non-vanishing value of either temperature or the transverse field have been considered in order to improve the success probabilities of quantum annealing
~\cite{2016PhRvE94c2105N,2017PhRvA96d2310N}.

In this section, we investigate whether \vrs~is capable of correcting such errors. 
One approach is to make the final Hamiltonian variational. 
Let us describe the accurate final Hamiltonian in terms of its components: $\hfin=\sum_{i} J_i \bf{\sigma}^i_{\text{fin}}$.
In the presence of control errors, the couplings of the final Hamiltonian on a quantum device have different values, which we denote by
\bea
\tilde{H}_{\text{fin}}=\sum_{i}\tilde{J}_i \bf{\sigma}^i_{\text{fin}}.
\eea
One way to use the variational method for mitigating systematic control errors is to treat the final Hamiltonian itself as variational.
Starting from $J^{(0)}_i = J_i$ as input couplings, one runs a time-dependent Hamiltonian
$H(t)=
A(t)\hini+ B(t)\tilde{H}_{\text{fin}}$, measures the expectation values of the Pauli words $\langle \bf{\sigma}^i_{\text{fin}} \rangle$, and then estimates the energy using the accurate coefficients
$E=\sum_i J_i  \langle \bf{\sigma}^i_{\text{fin}} \rangle$.
Based on the result, a classical optimizer updates the input couplings $J^{(k-1)}_i \to J^{(k)}_i$ so that the couplings on the  quantum device 
$\tilde{J}^{(k)}_i $ become closer to the desired accurate values $J_i$.

As an alternative method, we can use a non-adiabatic process to obtain accurate results.
In this case, we fix the final Hamiltonian, and adjust the scheduling during the annealing process so that the final quantum state becomes closer
to the exact ground state.
Let us consider the following time-dependent Hamiltonian in \vrs:
\bea
H(t)=
A(t)\hini+ B(t)\tilde{H}_{\text{fin}}+
\sum_{i} C_{i}(t) \bf{\sigma}^i_{\text{fin}}.
\label{inaccurate-H}
\eea
The last term in Eq.~(\ref{inaccurate-H}) is the navigator Hamiltonian of GUCC that was used in \cite{VanQver}. 
Note that in the \vrs~setting, one can absorb $B(t)$ into $C_i(t)$ by changing the boundary condition for $C_i(t)$,
as the Pauli words in the second term and the third term in Eq.~(\ref{inaccurate-H})
are the same. Here, we treat $B(t)$ and $C_i(t)$ separately for the purpose of analyzing the contributions of the navigator Hamiltonian and the final Hamiltonian.
As before, after running the time-dependent Hamiltonian,  the expectation values of each of the Pauli words $\langle \bf{\sigma}^i_{\text{fin}} \rangle$ are measured.
Then, the total energy is calculated by using the accurate coefficients
$E=\sum_{i} J_i \langle \bf{\sigma}^i_{\text{fin}} \rangle$. 
It is this quantity $E$ that an optimizer on a classical computer minimizes. 
This optimization process works as it is the ground state of $\hfin=\sum_{i} J_i \bf{\sigma}^i_{\text{fin}}$ that minimizes the function $\sum_{i} J_i \langle \bf{\sigma}^i_{\text{fin}} \rangle$.

\begin{figure}[ht]
\centering
\includegraphics[scale=0.4]{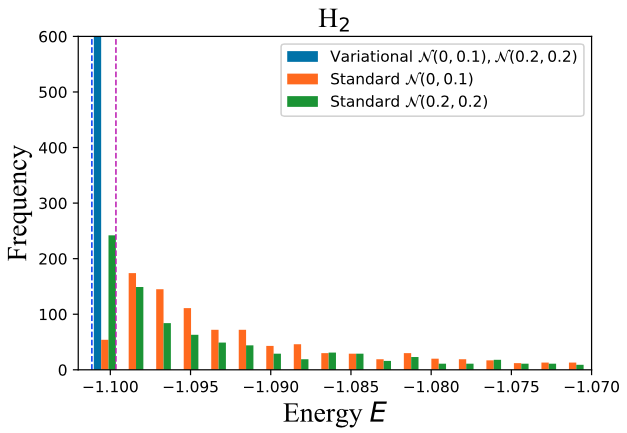}
\caption{
Frequency of the obtained energy of H$_\text{2}$, with \mbox{$d=1.0$~$\ang$} and an inaccurate final Hamiltonian.
The split number and group numbers are chosen to be
$(S,I,F)=(10,2,4)$. The annealing time is chosen to be $T=1$ for \vrs~and  $T=20$ for standard ASP.
}
\label{fig:Histgram-E-inaccurateH-QEC_GUCC}
\end{figure}

To demonstrate this method, we consider the hydrogen molecule. We add Gaussian noise to the coefficients 
\bea
 \tilde{J}_i = J_i + \xi_i\,,
\eea
where $\xi_i \in \mathcal{N}(\alpha,\beta)$, with $\alpha$ and $\beta$ being the mean and the standard deviation, respectively. 
We estimate the energy using different noise values $\xi_i$ for both standard ASP and \vrs.
A histogram of obtained energies is shown in Fig.~\ref{fig:Histgram-E-inaccurateH-QEC_GUCC}.
For both standard ASP and \vrs, the annealing time is chosen to be long enough to reach chemical accuracy when the accurate Hamiltonian $\hfin$ is used: $T=20$ for standard ASP
and $T=1$ for~\vrs.
Two distributions representing the control errors are considered; one with a zero mean $\mathcal{N}(0,0.1)$
and the other with a non-zero mean $\mathcal{N}(0.2,0.2)$.
As shown in Fig.~\ref{fig:Histgram-E-inaccurateH-QEC_GUCC}, the obtained energy in  standard ASP has a
wide range of distribution in $E$. This is expected, as standard ASP generates a ground state of the inaccurate Hamiltonian $\tilde{H}_{\text{fin}}=\sum_{i}\tilde{J}_i \bf{\sigma}_i$.
On the other hand, in~\vrs, the obtained energies always remain within  chemical accuracy.
Therefore, we conclude that \vrs~has resilience against control errors.

\subsection{Decoherence}

Another type of error which significantly impacts quantum computing is decoherence. In this section, we consider the impact of decoherence by solving the Lindblad master equation.
The introduction of noise into the system was achieved by adjusting the coefficients, \(C_{n}\) (Eq.~(\ref{eq: collapse})), of the Lindblad master equation, where the \(\sqrt{\gamma_{n}}\) term is representative of the noise strength, and \(A_{n}\) are collapse operators through which the noise is coupled to the system.
\begin{multline}
\dot{\rho}(t)=-\frac{i}{\hbar}[H(t), \rho(t)]+\\
\sum_{n} \frac{1}{2}\left[2 C_{n} \rho(t) C_{n}^{+}-\rho(t) C_{n}^{+} C_{n}-C_{n}^{+} C_{n} \rho(t)\right]\,,
\label{eq: master-eq}
\end{multline}
where
\begin{equation}
C_{n}=\sqrt{\gamma_{n}} A_{n}\,.
\label{eq: collapse}
\end{equation}
For simplicity, we consider  bit-flip noise.

\begin{figure}[ht]
\centering
\includegraphics[scale=0.46]{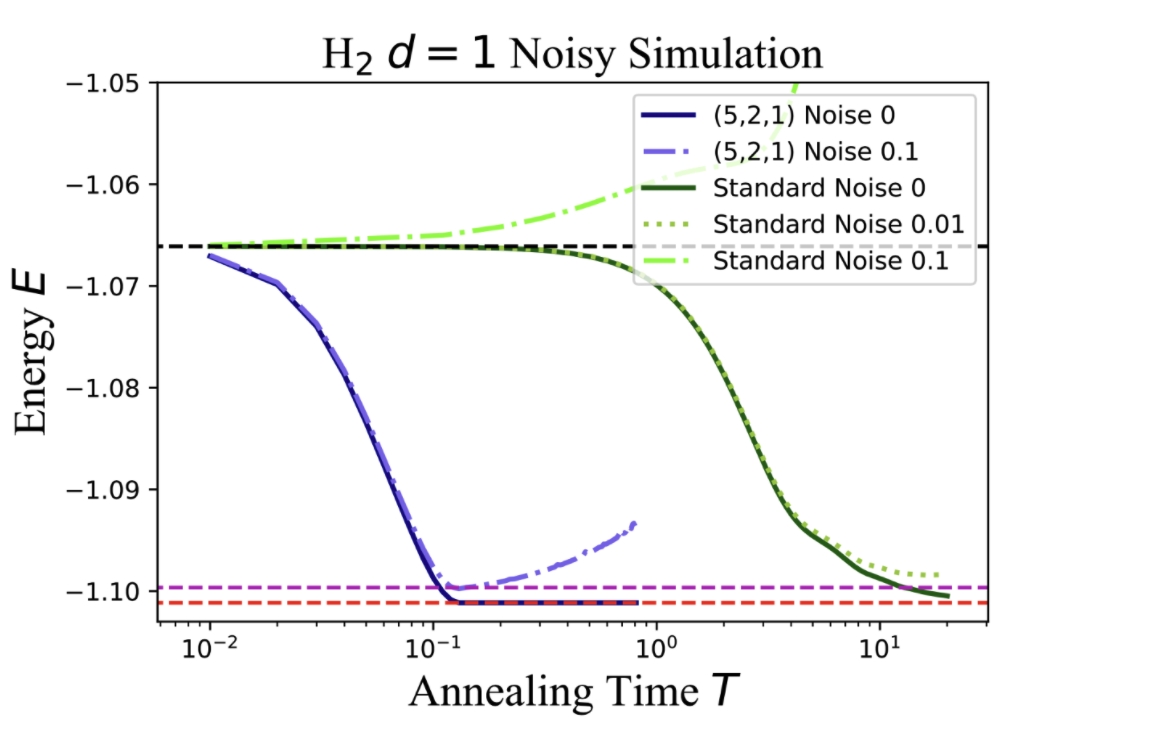}
\caption{
Results of VSQS as implemented to solve the Lindblad master equation (\ref{eq: master-eq}) {with bit-flip noise}, compared with those of standard ASP for an \(H_{2}\) system with a nuclear separation distance of \mbox{$d=1.0$~$\ang$}. These results compare the performance of these algorithms under three noise conditions: $noise~strength = 0$, $noise~strength = 0.01$, and $noise~strength = 0.1$.
The split number and group numbers are chosen to be
$(S,I,F)=(5,2,1)$. A logarithmic time scale has been employed on the horizontal axis of the figure in order to facilitate ease of visual comparison.
}
\label{fig:noise}
\end{figure}

Figure \ref{fig:noise} showcases the results obtained from VSQS as implemented to solve the Lindblad master equation, compared with those of standard ASP for a noisy \(H_{2}\) system with a nuclear separation distance of \mbox{$d=1.0$~$\ang$}. 
For this particular segment of the investigation, VSQS is set to allow a maximum amplitude of 10, a choice of optimizer of L-BFGS-B, a Hamiltonian split number of 5, and initial and final group numbers of 2 and 1, respectively. These results compare the performance of these algorithms under three noise conditions: $noise~strength = 0$, $noise~strength = 0.01$, and $noise~strength = 0.1$.

In the case where both routines are executed without any noise, VSQS (solid dark blue line) is observed to reach the \(H_{2}\) system’s exact energy approximately two orders of magnitude quicker than in the case of standard AQC (solid dark green line). Additionally, in the case where a noise strength of 0.1 has been introduced to both the VSQS and standard ASP systems, VSQS (dashed blue line) is observed to reach chemical accuracy quite rapidly, while standard ASP (light-green dashed line) fails to converge to chemical accuracy.

Even in the case where VSQS with $noise~strength = 0.1$ (dashed blue line) is compared with standard ASP $noise~strength = 0.01$ (small dotted green line), it is observed that VSQS outperforms standard ASP even with the noise in the VSQS system being greater by an order of magnitude, as VSQS is able to achieve chemical accuracy while standard ASP fails to converge to chemical accuracy altogether.

From this, we can conclude that, despite possessing additional control parameters, VSQS offers an obvious advantage over standard ASP in the context of annealing time required to achieve accurate results. Moreover, it achieves a level of robustness against the introduction of system noise when compared to the results obtained using standard ASP.


\section{Excited States}

Lastly, we briefly explain how to compute excited states by using VSQS.
There has been a lot of progress in calculating excited states on gate model quantum computers, especially in the NISQ era. Many of them are directly applicable to the VSQS algorithm. In this section, we take examples of folded spectrum techniques~\cite{PhysRev.46.828,doi:10.1063/1.466486,2014NatCo...5.4213P}
and
quantum subspace expansion (QSE)~\cite{2017PhRvA..95d2308M} and explain how they fit in the context of VSQS.

\subsection{Folded Spectrum}
In the folded spectrum technique, the cost function is replaced by
\bea
E(\mu)= \langle (H-\mu)^2 \rangle\,, 
\eea 
where $\mu$ is a parameter
\cite{PhysRev.46.828,doi:10.1063/1.466486,2014NatCo...5.4213P}. 
This technique is closely related to the variance minimization method, which is demonstrated in the context of quantum computing in Ref.~\cite{2019Natur.569..355K}. 
One drawback of this folded spectrum method is that the number of required measurements is increased significantly.
One option to mitigate this problem may be to use the classical shadow method with random single-qubit unitaries~\cite{2020NatPh..16.1050H,2020arXiv200615788H}.

\subsection{Quantum Subspace Expansion}
\label{Sec: Quantum Subspace Expansion}

In VSQS, the parametric AQC generates a quantum state $|\psi(\bf{a}, \bf{b}, \bf{c})\rangle $.
After optimizing the variational parameters, one obtains a  quantum state $|\psi_{GS}\rangle = |\psi(\bf{a}_{\min}, \bf{b}_{\min}, \bf{c}_{\min})\rangle$, which minimizes the expectation value of the Hamiltonian. 
This quantum state could be an accurate ground state or an approximate ground state. The latter could happen when the set annealing time is too short 
or when decoherence, unitary errors, or state preparation and measurement errors occur. Whether $|\psi(\bf{a}_{\min}, \bf{b}_{\min}, \bf{c}_{\min})\rangle$ is
exact or only approximately accurate, QSE can provide a way to calculate energies of excited states as well as
mitigate errors 
without additional AQC or single-qubit gate operations.
In QSE, a search space is expanded as
\bea
\left\{|\psi_{GS}\rangle , P_{\al}|\psi_{GS}\rangle  \right\}\,,
\eea
where $P_{\al}$ are some Pauli operators or linear combinations of them. There are various ways to choose $P_{\al}$:
one example is to choose a fermionic excitation operator such as $a^{\dagger}_{i}a_{j}$, and another example is to choose a single Pauli operator $\sigma^{k}_i$, where $k\in \{x,y,z\}$. In this expanded subspace,
the eigenvalue problem is formulated as
\bea
HC=SCE,
\label{eq: qse equation}
\eea
where 
the Hamiltonian $H$ is given by
\bea
H_{\al \beta}=\left\langle\psi_{GS}\left|P_{\al}^{\dagger} H P_{\beta}\right| \psi_{GS}\right\rangle
\eea
and the overlap matrix $S$ is given by 
\bea
S_{\al \beta} = 
\left\langle\psi_{GS}\left|P_{\al}^{\dagger}  P_{\beta}\right| \psi_{GS}\right\rangle.
\eea
$C$ represents the eigenstates in this subspace.
For a given state $|\psi_{GS}\rangle$ generated on a quantum computer, one can measure all the matrix elements of 
$H_{\al\beta}$ and $S_{\alpha\beta}$ simply by choosing the measurement axes $X, Y$ or $Z$. This is done by specifying the single-qubit rotations at the end of the annealing ($R_i(\phi_i)\in \text{Cl(2)}$ in Fig.~\ref{fig:InhomoVanQ loop}). Therefore, no additional quantum operations are needed for calculating excited states or to mitigate errors. The eigenvalue problem in Eq.~(\ref{eq: qse equation}) can be solved on a classical computer.

\section{Quantum Speedup and Shortcuts to Adiabaticity}
\label{Quantum Speedup and Shortcuts to Adiabaticity}

A considerable amount of interest has been generated towards achieving shortcuts to adiabaticity \cite{quan2010testing}. This itself is motivated by fundamental questions pertaining to the ``quantum limits'' experienced by processing speeds, and the overall viability of adiabatic computing as a successful computing regime.

Typically, quantum annealing is operated adiabatically, which means that the annealing time of the problem is much larger than the smallest energy gap between the ground state and the first excited state that is encountered in the evolution of the system.
However, it is also possible to consider QA operated in a non-adiabatic regime, where just as in the adiabatic case the goal remains to end the evolution of the system in the ground state of the final Hamiltonian. The difference in the non-adiabatic QA approach is that the system \textit{can} undergo diabatic transitions to excited states, after which it will return to the ground state.

A particular model for quantum computation that can be naturally thought of within the context of QA is that of "quantum walks", which is considered a viable prospect for achieving an advantage over classical algorithms, in the case of certain applications \cite{crosson2020prospects}. It has been demonstrated that any classical algorithm must necessarily take exponential time to solve this problem, yet a specific quantum walks--based algorithm known as \textit{glued trees}, originally introduced in Ref.~\cite{childs2003exponential}, was shown to solve the problem in polynomial time. Following this, a diabatic QA algorithm was presented by authors in Ref.~\cite{somma2012quantum}, which  was also shown to solve the problem in polynomial time. 

Regarding the work done in Ref.~\cite{somma2012quantum}, the QA evolution in the algorithm transitioned the system from the ground state to the first excited state, then back down to the ground state, which was enabled by the Hamiltonian spectrum. 
This realization of the glued trees problem is the only explicit example known to date for which a quantum annealing algorithm provides an exponential speedup compared to its classical counterpart, and it did so by partly using excited state evolution. As such, using the excited state approach employed in \cite{somma2012quantum} is likely the best hope for achieving a quantum advantage in this type of context. 

It is worth noting that while the special case of VSQS, namely, $I=F=1$, is certainly connected to the previous work done in \cite{somma2012quantum}, and could be thought of as a particular mapping of this problem, VSQS is, overall, a more general framework whose scope goes beyond the special $I=F=1$ Hamiltonian case.
Further details regarding the work presented in Ref.~\cite{somma2012quantum} can be found~\cite{crosson2020prospects}, which offers a review where the prospects for algorithms within the general framework of quantum annealing achieving a quantum speedup relative to classical state-of-the-art methods in combinatorial optimization and related sampling tasks.

Additionally, there exists a variety of other methods for shortcuts to adiabaticity (STA), some of which bear similarities to VSQS, such as the fast-forward approach. While some
of these techniques have similarities, instead of variationally determining the scheduling functions (as is the case with VSQS), the fast-forwarding approach instead modifies the Hamiltonians in a distinct way that involves variationally determining a potential term in order to generate a state close to an adiabatically prepared state without actually following the adiabatic path of the original Hamiltonians. Further details on the fast-forwarding approach, its similarities to and differences from other STA techniques, as well as a separate investigation of many other shortcut 
methods, are summarized in~\cite{RevModPhys.91.045001} and references therein.

\section{Conclusion}
\label{Sec. Conclusion}

In this paper, we proposed performing quantum simulations using variationally determined schedule functions. Our main objective was to shorten the individual runtimes during iterations of annealing, which is essential for obtaining accurate results in the NISQ era of quantum computing, which is marked by the absence of error correction techniques.
We conducted numerical simulations with the intent of investigating several characteristics of the \vrs~algorithm, and achieved this aim by employing VSQS to find the ground states of various molecular systems (such as H$_\text{2}$ and P4), as well as to solve an Ising model.
We analyzed the effects and advantages gained by the introduction of distinct schedule functions for distinct spins, as well as by optimizing the schedule functions variationally.
For computationally simple problems, such as finding the ground state of a hydrogen molecule, the optimization of the schedule functions enabled us to shorten the time of a single annealing run significantly.
For more challenging problems, namely the square configuration of the P4 molecule
 in which the highest occupied and lowest unoccupied molecular orbitals become degenerate, the use of distinct schedule functions for distinct terms was found to be the most influential contributing factor to shortening the annealing time. \vrs~allows quantum states to have an overlap with instantaneous excited states in a certain way during ASP, which results in achieving a significant overlap with the true ground state at the end of the annealing process.

We also applied the \vrs~algorithm for the triangular Ising model, which is understood to possess randomness and frustration. 
In this model, we introduced a navigator Hamiltonian consisting of $\sigma^{x}\sigma^{x}$ couplings. We found that the time-dependent Hamiltonian becomes non-stoquastic
for the optimal schedule functions.

Furthermore, we demonstrated the \vrs~algorithm's resilience to control errors. To do this, we considered the case where the final Hamiltonian was inaccurately implemented. 
When the quantum state followed the adiabatic path, then the final state became the ground state of the inaccurate final Hamiltonian. Therefore, the computational results provide inaccurate answers for  optimization problems.
The \vrs~algorithm allows for adjustments to be made to the schedule functions so that the final state minimizes the accurate Hamiltonian instead of the inaccurate Hamiltonian.
By doing so, we successfully generated the true ground state of a hydrogen molecule.

VSQS can be applied to any kind of implementation of quantum annealing. For instance, Ref.~\cite{PhysRevA.103.022619} has demonstrated VSQS in the Lechner--Hauke--Zoller scheme~\cite{Lechner:2015}.

\section{Acknowledgement}
We thank Marko Bucyk for reviewing and editing the manuscript.

\vfill\eject

\section*{Appendix A: Geometry of the P4 molecule}

\begin{figure}[ht]
\centering
\includegraphics[scale=0.37]{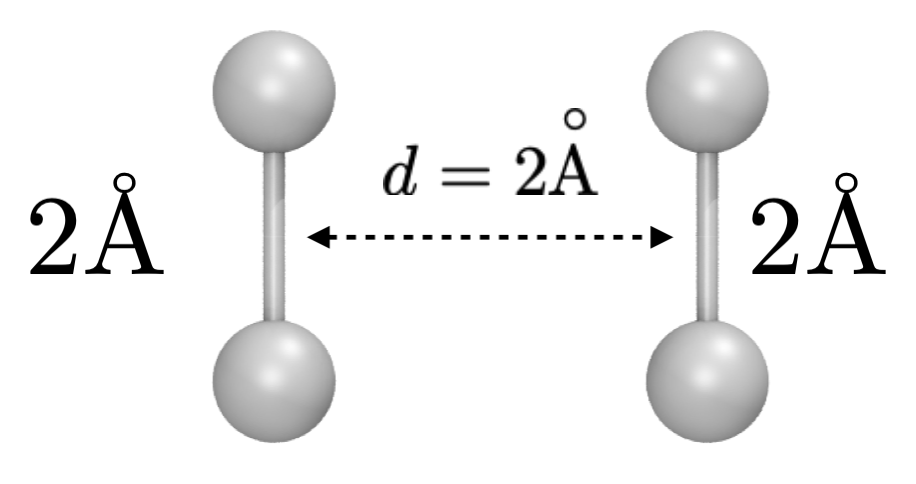}
\caption{
P4 molecule. The spheres represent hydrogen atoms.
}
\label{fig:P4mol}
\end{figure}

Figure~\ref{fig:P4mol} shows the geometry of the P4 molecule. The internuclear and intermolecular distances are chosen to be $2~\ang$.

\vfill
\section*{Appendix B: Standard ASP results for the hydrogen molecule}

\begin{figure}[ht]
\centering
\includegraphics[scale=0.37]{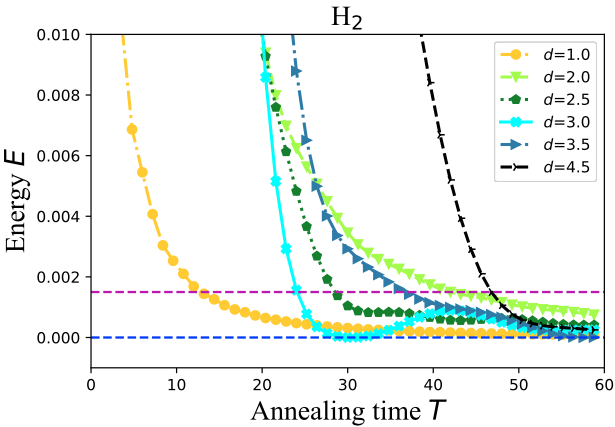}
\caption{
Energy $E$ versus time $T$ for the hydrogen molecule when using standard ASP
}
\label{fig:EvsT_standard_H2}
\end{figure}

The results of running standard ASP for the hydrogen molecule are shown in Fig.~\ref{fig:EvsT_standard_H2}.
The energy values exhibit non-monotonic behaviour as a function of the annealing time $T$ for
$d=2.5~\ang$ and $d=3.0~\ang$. This is evidence of coherent oscillation between different energy levels. See, for instance,  \cite{PhysRevA.88.013818,Messina_2014,
PhysRevA.95.032335,2019arXiv190300574M}.

\bibliography{ref6.bib}


\end{document}